\documentstyle[prl,aps,epsf,float,twocolumn]{revtex}

\begin{document}
\draft
\twocolumn[\hsize\textwidth\columnwidth\hsize\csname
@twocolumnfalse\endcsname

\title{
\hfill{\small{FZJ-IKP(TH)-2000-09}}\\[0.6cm]
The  pion charge radius from charged pion electroproduction}

\author{
V\'eronique~Bernard,$^1$ Norbert~Kaiser,$^2$ Ulf-G.~Mei{\ss}ner$^3$
}

\address{
$^{1}$Laboratoire de Physique Th\'eorique, 
Universit\'e Louis Pasteur, F-67084 Strasbourg Cedex, France \\
{Electronic address:~bernard@lpt6.u-strasbg.fr}\\
$^{2}$Physik Department, Technische Universit\"at M\"unchen, 
D-85747 Garching, Germany \\
{Electronic address:~Norbert.Kaiser@Physik.TU-Muenchen.de}\\
$^{3}$Institut f\"ur Kernphysik (Theorie), Forschungszentrum J\"ulich, 
D-52425 J\"ulich, Germany\\
{Electronic address:~Ulf-G.Meissner@fz-juelich.de} 
}
 
\maketitle

\begin{abstract}
We analyze a low--energy theorem of threshold pion electroproduction
which allows one to determine the charge radius of the pion. We show
that at the same order where the radius appears, pion loops induce a
correction to the momentum dependence of the longitudinal dipole 
amplitude $L_{0+}^{(-)}$. This model--independent correction amounts
to an increase of the pion charge radius squared from the electroproduction
data by about 0.26~fm$^2$. 
It sheds light on the apparent discrepancy between the recent
determination of the pion radius from electroproduction data and the
one based on pion--electron scattering.\\ $\,$
\end{abstract}

\vspace{0.8cm}

]

\noindent The charge (vector) radius of the pion is a fundamental quantity in
hadron physics. It can essentially be determined in two ways. One method is
pion scattering off electrons (or electron--positron
annihilation into pion pairs), this leads a pion root--mean--square
(rms) radius of\cite{Amen}
\begin{equation}
\langle r^2_\pi \rangle_V^{1/2} = (0.663 \pm 0.006)~{\rm fm}~,
\end{equation}
if one insists on the correct normalization of the pion charge (vector)
form factor, $F_\pi^V (0) =1$ (in units of the elementary charge $e$).
A more recent determination of the pion vector radius from low--momentum
space-- and time--like form factor data based on a very precise two--loop 
chiral perturbation theory representation\cite{BCT} gives 
\begin{equation}
\langle r^2_\pi \rangle_V^{1/2} = (0.661 \pm 0.012)~{\rm fm}~, 
\end{equation}
consistent with the value given above. This
number can be semi--quantitatively be understood in a naive vector
meson dominance picture, $\langle r^2_\pi \rangle_V^{1/2} =
\sqrt{6}/M_\rho \simeq 0.63\,$fm. The second method is based on
charged pion electroproduction, $\gamma^\star p \to \pi^+ n$. Here,
$\gamma^*,p,n$ and $\pi^+$ denote the virtual photon, the proton, the
neutron and the positively charged pion, in order. The unpolarized
cross section in parallel kinematics decomposes into a transversal
and a longitudinal piece.\footnote{For a textbook discussion on this
  issue, we refer the reader to Ref.\cite{AFF}.}
 While the former is sensitive to the the
nucleon axial radius, the latter is quite sensitive to the pion form
factor, i.e. to the pion radius for small momentum transfer. A recent 
measurement at the Mainz Microtron MAMI--II led to a pion radius
of\cite{MAMI}
\begin{equation}\label{MAMIrad}
\langle r^2_\pi \rangle_V^{1/2} = (0.74 \pm 0.03)~{\rm fm}~,
\end{equation}
which is a sizeably larger value than the one obtained from $\pi e$
scattering. It was hinted in Ref.\cite{MAMI} that their larger
value for the pion radius might be due to the inevitable
model--dependence based on the Born term approach to extract the
pion radius. It was also stated that there might be an additional
correction obtainable form chiral perturbation theory as it is the
case for the nucleon axial radius. Such a correction based on pion
loop diagrams had been predicted in 1992 for the axial
radius\cite{bkmax} and was verified by the MAMI experiment
to a good precision. A list of references pertaining to previous 
determinations of the nucleon axial radius from (anti)neutrino--proton
scattering and  pion electroproduction can be found in Ref.\cite{MAMI}.
 Here, we show
that there is indeed a similar kind of correction for the pion radius.
This new term modifies the momentum dependence of the longitudinal
S--wave amplitude $L_{0+}^{(-)}$ and leads one to expect an
even larger pion charge radius than the one given in
Eq.(\ref{MAMIrad}). It is conceivable that higher order corrections
yet to be calculated or contributions from higher multipoles
will completely resolve the discrepancy between the pion radius
determined from $\pi e$ scattering on one side and from charged pion
electroproduction on the other.

\medskip \noindent
Chiral perturbation theory allows one to make model--independent
statements based solely on the spontaneously and explicitly broken
chiral symmetry of QCD. S--matrix elements and transition currents are
systematically expanded in external momenta and quark (meson) mass
insertions, collectively denoted by a small parameter $q$. Based on
the underlying power counting, to a given order, one has to consider
tree as well as pion loop graphs. Of particular interest are the so--called
low--energy theorems, which give predictions to a certain order
expressed entirely in terms of measurable quantities (a detailed
discussion about this issue can be found in Ref.\cite{EM}).
Our starting point is the low--energy theorem for the longitudinal
dipole amplitude $L_{0+}^{(-)}$, accessible in charged pion
electroproduction.\footnote{We use standard notation for the multipoles, 
the superscript
refers to the isospin, the subscripts are $l\pm$, with $l$ the orbital angular
momentum of the pion--nucleon system and the total angular momentum 
is $j = l \pm 1/2$. For a detailed discussion of the kinematics and
multipole decomposition in pion electroproduction, see refs.\cite{AFF,bkmpr}.}
It has been derived from baryon chiral perturbation
theory to third order in the chiral expansion and reads\cite{bkmpr}:
\begin{eqnarray}
L_{0+}^{(-)} (\mu , \nu)  &=&  E_{0+}^{(-)} (\mu , \nu) \nonumber \\
&+& {e g_{\pi N}
  \over 8\pi m }\,
(\mu^2 -\nu) \biggl\{ {\kappa_v \over 4} + {m^2\over 6} \langle r_A^2
\rangle  \nonumber \\
&+& {\sqrt{(2+\mu)^2 -\nu} \over 2(1+\mu)^{3/2} (\nu - 2\mu^2
  -\mu^3)} \nonumber \\
&+& \biggl( {1\over \nu -2\mu^2} - {1\over \nu} \biggr)
(F_\pi^V (m^2\nu) -1) \nonumber \\
&+&  {m^2\over 8\pi^2 F_\pi^2}\Xi_4(-\nu\mu^{-2}) 
\biggr\} + {\cal O}(q^3)~, \\
E_{0+}^{(-)} (\mu , \nu) &=& {e g_{\pi N}  \over 8\pi m }\,
  \biggl\{ 1 - \mu +C\mu^2 \nonumber \\
&+& \nu \biggl(  {\kappa_v \over 4} + {1\over
  8} + {m^2\over 6} \langle r_A^2\rangle \biggr) \nonumber \\
&+& {\mu^2 m^2 \over
  8\pi^2 F_\pi^2} \Xi_3(-\nu\mu^{-2}) \biggr\} + {\cal O}(q^3)~,  
\end{eqnarray}
with
\begin{eqnarray}
\Xi_3 (\rho) & = & \sqrt{1+{4\over \rho}} \ln \biggl(\sqrt{1+{\rho
  \over 4}} + {\sqrt{\rho} \over 2} \biggr)  \nonumber \\
   &+& 2 \int_0^1 \sqrt{(1-x) [
  1 + x(1+\rho )] }  \nonumber \\
  &\times& \arctan {x\over \sqrt{(1-x)[1+x(1+\rho)]} }~, \\
\Xi_4 (\rho) & = & \int_0^1 dx {x(1-2x)\over
  \sqrt{(1-x)[1+x(1+\rho)]}} \nonumber \\
   &\times&\arctan {x\over \sqrt{(1-x)[1+x(1+\rho)]
  }}~,
\end{eqnarray}
in terms of the dimensionless quantities $\mu = M_\pi/m$ and $\nu =
k^2/m^2$, with $M_\pi$ $(m)$ the charged pion (nucleon) mass and
$k^2 \le 0$ the photon virtuality.  Furthermore, $\kappa_v = \kappa_p -
\kappa_n = 3.71$ is the isovector anomalous moment of the nucleon,
$F_\pi = 92.4$~MeV the weak pion decay constant, $g_{\pi N} = 13.4$ 
the pion--nucleon coupling constant and
$C \simeq 0.4$ depends on some low--energy constants. Its precise
form is not of interest here but can be found in
Ref.\cite{bkmpr}. We remark that these expressions have been obtained
in a relativistic version of baryon chiral perturbation theory and that
some of the kinematical prefactors have not been expanded. However,
since we are only after a leading order pion loop effect, the result
will not be different from a heavy baryon analysis~\cite{jm,bkkm}. 
For calculating
higher order corrections consistently, one either has to use the
heavy baryon formalism or the recently proposed 
infrared regularization~\cite{BL}.
Note also that the dependence of $L_{0+}^{(-)}$ on
the axial radius is very weak because the terms $\sim \nu \langle
r_A^2\rangle$ cancel (that is why the axial radius discrepancy
discussed before resides in the electric dipole amplitude\cite{bkmax}). 
As announced, the pion vector form factor $F_\pi^V (k^2)$ appears in the
expression for the longitudinal multipole.  The pion form factor has
the following low--energy expansion,
\begin{equation}
F_\pi^V (k^2) = 1 + {1\over 6} \langle r^2_\pi \rangle_V \, k^2 +
{\cal O}(k^4)~.
\end{equation}
Consequently, to separate the pion radius, one should consider the
slope of the longitudinal multipole. For doing that, one has to expand
the functions $\Xi_{3,4} (-\nu\mu^{-2})$ in powers of $k^2 = \nu m^2$
and pick up all terms proportional to $\nu$. This gives:
\begin{eqnarray}\label{L0}
{\partial L_{0+}^{(-)} \over \partial k^2} \biggl|_{k^2 = 0} &=&
{e g_{\pi N} \over 32 \pi m} \biggl\{ {1 \over M_\pi^2} - {1\over
  mM_\pi} + {13\over 8m^2} \nonumber\\
+ {1\over 3} \langle r_\pi^2 \rangle_V &+&
{1\over 32 F_\pi^2} \biggl( {16\over \pi^2} -1 \biggr)
+ {\cal O} (M_\pi) \biggr\}~.
\end{eqnarray}
The first four terms are standard~\cite{scherer}, they comprise the 
conventional dependence on the pion vector radius, recoil effects and the
dominant chiral limit behavior of the slope of the longitudinal
multipole. The strong $1/M_\pi^2$ chiral singularity stems from the
$k^2$--derivative of the pion--pole term which appears already at leading order
in the chiral expansion. 
\begin{figure}[htb]
\begin{center}
\hspace{0.8cm}
\epsfxsize=1.9in
\epsffile{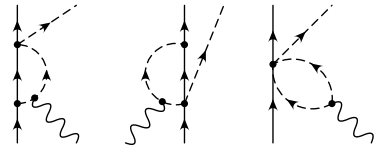}
\vspace{0.3cm}
\caption{Relevant one--loop  diagrams. Crossed partners are not shown.
The solid, dashed and wiggly lines denote nucleons, pions and photons,
in order.\label{fig:1}}
\end{center}
\end{figure}     
\noindent
The last term in Eq.(\ref{L0}) originates from the so--called
triangle and tadpole (with three pions coupling to the nucleon at one
point) diagrams which are known to play a prominent role in pion photo--
and electroproduction (see Fig.~\ref{fig:1}). 
The formal reason for the appearance of 
this new, model--independent contribution at order $k^2$ is that one cannot
interchange the order of taking the derivative at $k^2 =0$ and the chiral limit
$M_\pi \to 0$. 
Consequently, all determinations of the pion radius from
electroproduction (based on tree-level amplitudes including nucleon and pion
form factors) have ``measured'' the modified radius,
\begin{equation}\label{modrad}
 \langle \widetilde{r}^2_\pi \rangle_V =  \langle r^2_\pi \rangle_V
+ {3\over 32 F_\pi^2} \biggl( {16\over \pi^2} -1 \biggr)~.
\end{equation}
The novel term on the right--hand-side of Eq.(\ref{modrad}) amounts
to 0.266~fm$^2$, a bit more than half of the squared pion rms radius,
$\langle r_\pi^2 \rangle_V \simeq 0.44\,$fm$^2$. Therefore, from the
longitudinal multipole alone, one expects to find a larger
pion radius if one analyses pion electroproduction based on Born
terms, 
\begin{equation}
 \langle \widetilde{r}^2_\pi \rangle_V = (0.44+0.26)~{\rm fm}^2 =
(0.83~{\rm fm})^2~,
\end{equation}
which is even larger than the result of the Mainz analysis, 
cf. Eq.(\ref{MAMIrad}). We point out, however, that the 
contribution of the pion radius to the derivative of the longitudinal
multipole is a factor of ten smaller than the one from the first
three terms in the curly brackets in Eq.(\ref{L0}). Therefore,
a fourth order analysis is certainly needed to further quantify
the ``pion radius discrepancy''. Furthermore, the pion form factor
contribution to the longitudinal cross section is also present in higher
multipoles. In fact, it is known that the convergence of the multipole
series for the pion pole term is slow. One should therefore also 
investigate such effects for these higher multipoles or directly compare the
predictions of complete one--loop calculation with the data of the
longitudinal electroproduction cross section.   For the purpose of
demonstrating the significance of chiral loop effects the $k^2$-slopes 
(considered here in Eq.(9) and in Ref.\cite{bkmax} for $E_{0+}^{(-)}$)  
are, however, best suited.  
What we have shown here is that as in the case of the nucleon axial 
mean square radius, the pion loops, which are a unique consequence of the
chiral symmetry of QCD, modify the naive Born term analysis and should be
taken into account.

\pagebreak

\bigskip

\end{document}